\documentclass[aps,prc,amsmath,superscriptaddress,amssymb,floats,nofootinbib]{revtex4-1}
\usepackage[pdftex]{graphicx}
\usepackage{setspace}
\oddsidemargin=\evensidemargin
\begin{document}  

\title{
{\large  \bf Studying the Proton ``Radius'' Puzzle with \boldmath$\mu p$
  Elastic Scattering} \\
{The MUon proton Scattering Experiment (MUSE) Collaboration}
}

\vspace{0.3cm}  
\author{R.~Gilman (Contact person)} 
\affiliation{Rutgers University, New Brunswick, New Jersey, USA}
\author{E.J.~Downie (Spokesperson)}
\affiliation{George Washington University, Washington, DC, USA} 
\author{G.~Ron (Spokesperson)}
\affiliation{Hebrew University of Jerusalem, Jerusalem, Israel}
\author{A.~Afanasev}
\affiliation{George Washington University, Washington, DC, USA}
\author{J.~Arrington}
\affiliation{Argonne National Lab, Argonne, IL, USA} 
\author{O.~Ates}
\affiliation{Hampton University, Hampton, Virginia, USA}
\author{F.~Benmokhtar}
\affiliation{Duquesne University, Pittsburgh, PA, USA} 
\author{J.~Bernauer}
\affiliation{Massachusetts Institute of Technology, Cambridge,
  Massachusetts, USA}
\author{E.~Brash}
\affiliation{Christopher Newport University, Newport News, Virginia, USA} 
\author{W.~J.~Briscoe}
\affiliation{George Washington University, Washington, DC, USA} 
\author{K.~Deiters}
\affiliation{Paul Scherrer Institut, CH-5232 Villigen, Switzerland}
\author{J.~Diefenbach}
\affiliation{Hampton University, Hampton, Virginia, USA}
\author{C.~Djalali}
\affiliation{University of Iowa, Iowa City, Iowa, USA}
\author{B.~Dongwi}
\affiliation{Hampton University, Hampton, Virginia, USA}
\author{L.~El Fassi} 
\affiliation{Rutgers University, New Brunswick, New Jersey, USA}
\author{S.~Gilad}
\affiliation{Massachusetts Institute of Technology, Cambridge,
  Massachusetts, USA}
\author{K.~Gnanvo}
\affiliation{University of Virginia, Charlottesville,  Virginia, USA}
\author{R.~Gothe}
\affiliation{University of South Carolina, Columbia, South Carolina, USA}
\author{D.~Higinbotham}   
\affiliation{Jefferson Lab, Newport News, Viginia, USA}
\author{R.~Holt}   
\affiliation{Argonne National Lab, Argonne, IL, USA}
\author{Y.~Ilieva}
\affiliation{University of South Carolina, Columbia, South Carolina, USA}
\author{H.~Jiang}
\affiliation{University of South Carolina, Columbia, South Carolina, USA}
\author{M.~Kohl}
\affiliation{Hampton University, Hampton, Virginia, USA}
\author{G.~Kumbartzki} 
\affiliation{Rutgers University, New Brunswick, New Jersey, USA}
\author{J.~Lichtenstadt}
\affiliation{Tel Aviv University, Tel Aviv, Israel}
\author{A.~Liyanage}
\affiliation{Hampton University, Hampton, Virginia, USA}
\author{N.~Liyanage}
\affiliation{University of Virginia, Charlottesville,  Virginia, USA}
\author{M.~Meziane}
\affiliation{Duke University, Durham, North Carolina, USA}
\author{Z.-E.~Meziani}
\affiliation{Temple University, Philadelphia, Pennsylvania, USA}
\author{D.G.~Middleton}
\affiliation{Institut f\"{u}r Kernphysik, Johannes Gutenberg Universit\"{a}t, Mainz 55099, Germany}
\author{P.~Monaghan}
\affiliation{Hampton University, Hampton, Virginia, USA}
\author{K.~E.~Myers}
\affiliation{Rutgers University, New Brunswick, New Jersey, USA}
\author{C.~Perdrisat}
\affiliation{College of William \& Mary, Williamsburg, Virginia, USA}
\author{E.~Piasetzsky}
\affiliation{Tel Aviv University, Tel Aviv, Israel}
\author{V.~Punjabi}
\affiliation{Norfolk State University, Norfolk, Virginia, USA}
\author{R.~Ransome}
\affiliation{Rutgers University, New Brunswick, New Jersey, USA}
\author{D.~Reggiani}
\affiliation{Paul Scherrer Institut, CH-5232 Villigen, Switzerland}
\author{P.~Reimer}   
\affiliation{Argonne National Lab, Argonne, IL, USA}
\author{A.~Richter}
\affiliation{Technical University of Darmstadt, Darmstadt, Germany}
\author{A.~Sarty}
\affiliation{St.~Mary's University, Halifax, Nova Scotia, Canada}
\author{E.~Schulte}
\affiliation{Temple University, Philadelphia, Pennsylvania, USA}
\author{ Y.~Shamai}
\affiliation{Soreq Nuclear Research Center, Israel}
\author{N.~Sparveris}
\affiliation{Temple University, Philadelphia, Pennsylvania, USA}
\author{S.~Strauch}
\affiliation{University of South Carolina, Columbia, South Carolina, USA}
\author{V.~Sulkosky}
\affiliation{Massachusetts Institute of Technology, Cambridge,
  Massachusetts, USA}
\author{A.S.~Tadepalli}
\affiliation{Rutgers University, New Brunswick, New Jersey, USA}
 \author{M.~Taragin}
\affiliation{Weizmann Institute, Rehovot, Israel}
\author{L.~Weinstein}
\affiliation{Old Dominion University, Norfolk, Virginia, USA}

\begin{abstract}
The Proton Radius Puzzle is the inconsistency between the proton
radius determined from muonic hydrogen and the proton radius determined
from atomic hydrogen level transitions and $ep$ elastic scattering.
No generally accepted resolution to the Puzzle has been found.
Possible solutions generally fall into one of three categories:
the two radii are different due to novel beyond-standard-model physics,
the two radii are different due to novel aspects of nucleon structure, and
the two radii are the same, but there are underestimated uncertainties
or other issues in the $ep$ experiments.

The MUon proton Scattering Experiment (MUSE) at the Paul Scherrer Institut
is a simultaneous measurement of $\mu^+ p$ and $e^+ p$ 
elastic scattering, as well as  $\mu^- p$ and $e^- p$ elastic scattering,
which will allow a determination of the consistency of the $\mu p$ 
and the $ep$ interactions. 
The differences between $+$ and $-$ charge scattering are sensitive to
two-photon exchange effects, higher-order corrections to the
scattering process.
The slopes of the cross sections as $Q^2 \to 0$ determine the proton ``radius''.
We plan to measure relative cross sections at a typical level of 
a few tenths of a percent, which should allow the proton radius to
be determined at the level of $\approx 0.01$~fm, similar to previous
$ep$ measurements.
The measurements will test several possible explanations of the
proton radius puzzle, including some models of beyond-standard-model
physics, some models of novel hadronic physics, and some issues
in the radius extraction from scattering data.

\end{abstract}

\maketitle
\tableofcontents   

\newpage
\begin{spacing}{2}

\section{Physics Motivation}

\subsection{Introduction}

The {\em Proton Radius Puzzle} 
refers to the disagreement between the proton charge radius
determined from muonic hydrogen and determined from
electron-proton systems: atomic hydrogen and $ep$ elastic scattering.
Up until 2010, the accepted value for the proton radius was 
$0.8768 \pm 0.0069$~fm, determined essentially from atomic
hydrogen measurements in the 2006 CODATA analysis \cite{Mohr:2008fa}.
The best $ep$ scattering result was probably 0.895 $\pm$ 0.018 fm, 
from the analysis of Sick \cite{Sick:2003gm}. 
The consistency of these two results made the muonic hydrogen
determination of $0.84184 \pm 0.00067$~fm by Pohl {\it et al.}
\cite{Pohl:2010zza} quite surprising.
The $\approx$5$\sigma$ discrepancy, in terms of the order of magnitude
less precise electron measurements, has attracted much attention.
It has motivated numerous invited talks, 
dedicated sessions at several meetings,
a Workshop on the Proton Radius Puzzle at the European
Center of Theory in Trento, Italy \cite{prpw:2012}, 
a review paper \cite{Pohlreview:2012},
some new experiments,
and stories in the popular media.
The paper by Pohl {\it et al.} has been cited 
about 200 times to date.
Some of the numerous suggestions for how the Puzzle  
might be resolved are discussed below.

\begin{figure}[b]
\centerline{\includegraphics[width=3.2in]{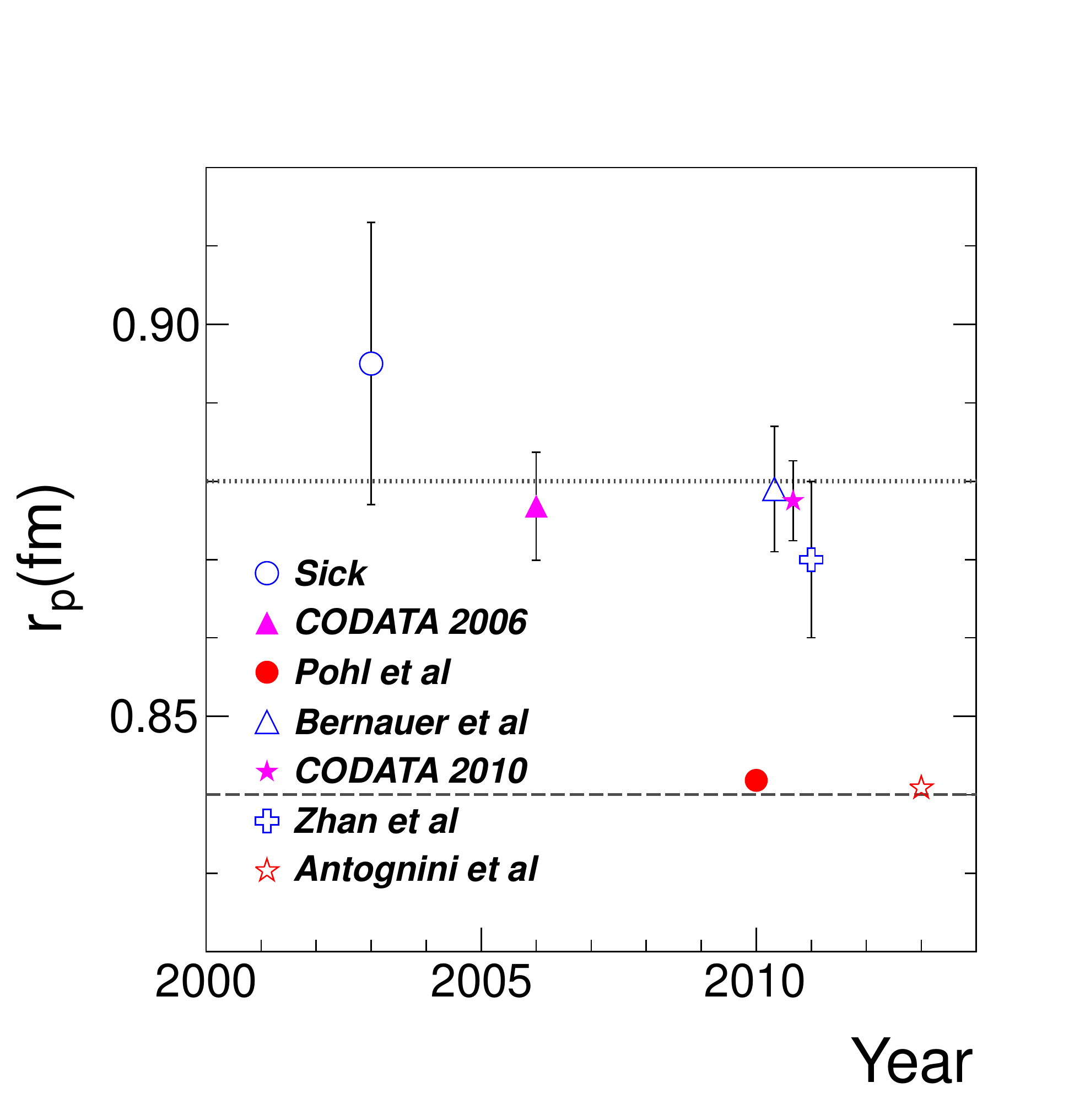}}
\caption{A summary of some recent proton charge radius
  determinations: Sick \cite{Sick:2003gm}, 
CODATA 2006 \cite{Mohr:2008fa}, 
Pohl {\it et al.} \cite{Pohl:2010zza},
Bernauer {\it el al.} \cite{Bernauer:2010wm}, 
CODATA 2010 \cite{Mohr:2010}, 
Zhan {\it et al.} \cite{Zhan:2011ji},
and
Antognini {\it el al.} \cite{Antognini25012013}.
}
\label{fig:radius}
\end{figure}

The Puzzle has been reinforced by three more recent  experimental results and
the 2010 CODATA analysis.
First, a precise $ep$ scattering cross section measurement
\cite{Bernauer:2010wm} at Mainz determined $\approx$1400 cross sections in the range 
$Q^2$ = 0.0038 $\to$ 1 GeV$^2$.
The Mainz analysis of only their data with a wide range of functional
forms led to a proton electric radius of 0.879 $\pm$ 0.008 fm.  
Second, an experiment \cite{Zhan:2011ji} at Jefferson
Lab measured $\vec{e}p \to e^{\prime}\vec{p}$ to determine 1\% form
factor ratios in the range Q$^2$ = 0.3 $\to$ 0.8 GeV$^2$.
An analysis of world data (excluding the Mainz data set but including
the data analyzed in \cite{Sick:2003gm}) resulted in a
radius of 0.870 $\pm$ 0.010 fm, consistent
with the Mainz electric radius determination  -- although there were
differences in the magnetic radius determination.
Third, a new muonic hydrogen measurement by Antognini {\it et al.} 
\cite{Antognini25012013} has recently reported a value for the proton
radius,  $r_p = 0.84087 \pm 0.00039$~fm,
in agreement with the Pohl {\it et al.} measurement.
Antognini {\it et al.} also report a magnetic radius consistent with
electron scattering results, though in this case with uncertainties a
few times larger. 
The 2010 CODATA analysis \cite{Mohr:2010} included the Mainz result --
the JLab result appeared too late to be included -- and adopted a
proton radius value of $r_p = 0.8775 \pm 0.0051$~fm.
The CODATA analysis concluded that:
{\em ``Although the uncertainty of the muonic hydrogen value is significantly
smaller than the uncertainties of these other values, its negative
impact on the internal consistency of the theoretically predicted and
experimentally measured frequencies, as well as on the value of the
Rydberg constant, was deemed so severe that the only recourse was to
not include it in the final least-squares adjustment on which the 2010
recommended values are based.''}
The Particle Data Group recently concluded that:
{\em ``Until the difference between the $ep$ and $\mu p$ values is understood,
it does not make sense to average all the values together. For the present,
we stick with the less precise (and provisionally suspect) CODATA
2012\footnote{Note that the CODATA 2010 result appeared in 2012.}
value. It is up to workers in this field to solve this puzzle.''}
Thus, the discrepancy between muonic and electronic measurements
of the proton radius has increased from 5$\sigma$ to 7$\sigma$
in the past almost 3 years, and the inconsistency of the results is
widely recognized.
A partial summary of recent proton radius extractions is shown in
Fig.~\ref{fig:radius}.

Arguably, the Proton Radius Puzzle is more puzzling today than when 
it first appeared.
Not only has the discrepancy increased, but numerous possible
explanations of the Puzzle have been shown to not work.
There have been suggestions of issues in the $\mu p$
radius determination, issues in the $ep$ radius determination,
novel hadronic physics, and novel beyond standard model (BSM) physics.
We briefly review the suggested explanations here.
More detail can be found in talks at the Trento Workshop \cite{prpw:2012}, 
and in the review paper by Pohl, Gilman, Miller and Pachucki
\cite{Pohlreview:2012}.

The finite size of the proton causes a small perturbation to the
Coulomb potential that basically shifts the energies of only $s$
states. The effect can be determined through Lamb shift measurements,
given a sufficiently accurate relativistic theory that accounts for
recoil corrections, vacuum polarization, etc., as the finite size
effect is rather small. The atomic physics calculations have now
all been repeated and verified by independent groups, and it is
believed that at the level of the muonic hydrogen experiment there
is no significant missing or uncalculated higher order physics.
The extraction of the radius from muonic hydrogen also requires
some knowledge of additional details of the proton's structure --
e.g., the third Zemach moment -- but it is generally believed that
there are no significant issues here; we will return to this point
below.
Experimentally, once the laser system exists, the muonic hydrogen
measurement is 8,000,000 times more sensitive to the proton
radius than an electronic hydrogen measurement, as $\psi(r=0) \propto
m^3_l$,
so the muonic hydrogen experiment appears to be the most solid 
of all the experimental results.

Issues in the $ep$ experiments would appear to be unlikely.
It would be odd if two independent techniques, atomic hydrogen
and $ep$ scattering, gave the same wrong result, especially as there
are two independent $ep$ analyses from different data sets.
However, the CODATA analyses neglect that the
atomic hydrogen measurements were done by only a few groups,
and thus likely there are some correlations between the results;
they are not entirely independent. Also, nearly all the atomic
hydrogen results are individually within 1$\sigma$ of the muonic
hydrogen result. Only one is 3$\sigma$ away. Only when all the
atomic hydrogen results are averaged does the discrepancy become
so impressive. Thus, the uncertainty in the atomic hydrogen
result is probably underestimated.

Numerous mistakes have been made over the years in determining the
radius from $ep$ scattering analyses, and there continues to be a
range of results.
The analysis of Sick \cite{Sick:2003gm} was arguably the first to
include all necessary ingredients to get a reliable answer, and more
recent analyses tend to as well, although typically insufficient 
attention is paid to the issue of model dependence.
In addition to the results reported by the experimenters above, 
we can consider the dispersion relation analysis of \cite{Lorenz:2012tm}
($r_p$ = 0.84 $\pm$ 0.01 fm with $\chi^2$/d.o.f $\approx$ 2.2),
the $z$ expansion of \cite{Hill:2010yb}
($r_p$ = 0.871 fm $\pm$ 0.009 fm $\pm$ 0.002 fm $\pm$ 0.002 fm),
the sum-of-Gaussians fit of \cite{Sick:2011zz,Sick:2012zz}
($r_p$ = 0.886 fm $\pm$ 0.008 fm),
and unpublished Taylor expansion fits to the low $Q^2$ data by
C.E.~Carlson and K.~Griffioen 
($r_p$ $\approx$ 0.84 fm).
Of these recent analyses, there are reasons to favor the two analyses 
yielding larger radii -- see \cite{Pohlreview:2012} --
but the variation in results does suggest that the uncertainty
arising out of the fits is underestimated.

There have been a number of suggestions of novel hadronic physics,
but almost none of them are accepted by experts as reasonable.
It is hard to see how narrow structures in the form factors or
anomalously large third Zemach moments arise out of conventional
hadronic physics.
The one existing viable idea \cite{Miller:2012ne} is that the uncertainty in the two-photon
exchange term coming from the proton polarizibility is underestimated;
changes in this term affect the radius extracted from muonic hydrogen.
Technically, evaluating the polarizibility requires elastic,
inelastic, and subtraction terms, where the subtraction term is needed
for convergence. 
The subtraction term diverges without the introduction of a form
factor, which has known behavior at small and large $Q^2$, but at
present does not appear constrained at intermediate $Q^2$.
Typical assumptions lead to the subtraction term contribution and
uncertainty having an effect that is only a few percent
of the Puzzle, but at present it appears that there is no constraint
from data -- only theoretical bias -- that prevents it from being much
larger.
We note that this explanation of the puzzle affects mainly the muon,
as the effect is proportional to the $m^4_{lepton}$, and that this
effect predicts enhanced two-photon exchange effects in muon
scattering from the proton.

If the experiments are not wrong, and there is no novel hadronic
physics, novel BSM physics has to be considered.
Previous measurements of lepton universality and numerous
other data, such as the muon $(g-2)$ measurements, 
constrain possible models of new physics.
Nevertheless, several models have been created.
Tucker-Smith and Yavin \cite{Tucker-Smith:2011}
found that a new scalar force carrier in the MeV mass range
is not ruled out by other data and could account for the
Proton Radius Puzzle. The main constraint is that the scalar
needs to have smaller coupling to the neutron than to the proton.
Batell, McKeen, and Pospelov \cite{Batell:2011qq} 
indicate that there are a number of ways new forces can evade
existing constraints but lead to the Proton Radius Puzzle.
In particular, they consider a combination of new vector and
scalar particles with masses of 10's of MeV.
The combination of two new particles allows the Puzzle to
be explained while evading other constraints.
This model leads to enhanced parity violation in muon scattering
and in muonic atom radiative capture.
Rislow and Carlson \cite{CarlsonRislow:2012:NewPhysisc}
show that one can explain the Puzzle while evading other constraints
by a combination of new scalar and pseudoscalar, or new
vector and pseudovector, particles.
The allowed coupling constants are constrained by the Puzzle and
muon $(g-2)$, and the mass ranges are constrained by $K$ decays,
but not too much if the new forces couple much more strongly
to muons than to electrons.
Thus there are a variety of possible BSM explanations
of the Puzzle, with parameters constrained by existing data, and
with potentially observable consequences in several 
experiments.

The various explanations of the Puzzle were
reviewed during the Proton Radius
Puzzle Workshop \cite{prpw:2012} in Trento, Italy from Oct 29 - Nov 2,
2012.  The workshop, organized by R. Pohl, G. A. Miller, and
R. Gilman, included nearly 50 experts in atomic and nuclear theory and 
experiment, as well as BSM theory.  At the end of the workshop, a vote
was held regarding the likely resolution of the Puzzle. The about equally
favored alternatives were BSM physics and issues in the $ep$
experiments. There was also support for the proton polarizibility
explanation described above, and a significant fraction of the
community that was uncertain about the most likely explanation.

A number of experiments that might help resolve the Puzzle
were discussed at the Workshop. Efforts to perform new
atomic hydrogen experiments in the next 5 - 10 years could
help confirm the Puzzle exists, or instead indicate consistency
in the muonic and electronic atomic physics measurements. 
A new muonic deuterium
experiment can be compared with the electron-deuteron
radius measurements to check for consistency.
A new Jefferson Lab experiment \cite{gasparian2011} 
approved by PAC39 plans to measure very low $Q^2$
electron scattering, from $\approx$ $10^{-4}$ GeV$^2$ to  $10^{-2}$ GeV$^2$,
perhaps as early as 2015.
We quote from the Jefferson Lab PAC:
``{\em Testing of this result is among the most timely and important
measurements in physics.}''
The efforts of the MUSE collaboration -- the focus of this White Paper -- to compare
$\mu^{\pm} p$ and $e^{\pm} p$ elastic scattering
were also discussed.
The Workshop conferees strongly supported all of
the experimental efforts; since the origin of the Puzzle is uncertain,
it is not clear which of the possible experiments will give us the data
that resolves the Puzzle.

To summarize, in the nearly 3 years since it appeared, the Proton
Radius Puzzle has become more puzzling, not less.
New experimental results confirm the puzzle.
Theoretical studies have ruled out many possible explanations,
leaving only a few possible.
The Puzzle has attracted wide interest, not just 
in the atomic, nuclear, and particle physics communities, but in the
popular science media as well, demonstrating the timeliness of resolving this issue.

\subsection{Muon-Proton Scattering Experiments}

The MUSE experiment was created on recognizing that the proton radius
has been measured in muonic and electronic atomic systems, and in
electron-proton elastic scattering, but not in muon-proton elastic
scattering.
Here we describe some previous tests of lepton universality, the
equivalence of muons and electrons, that were largely done about 
30 years ago.
We will focus on $\mu p$ and $ep$ scattering.

\begin{figure}[hb]
\centerline{\includegraphics[width=3.2in]{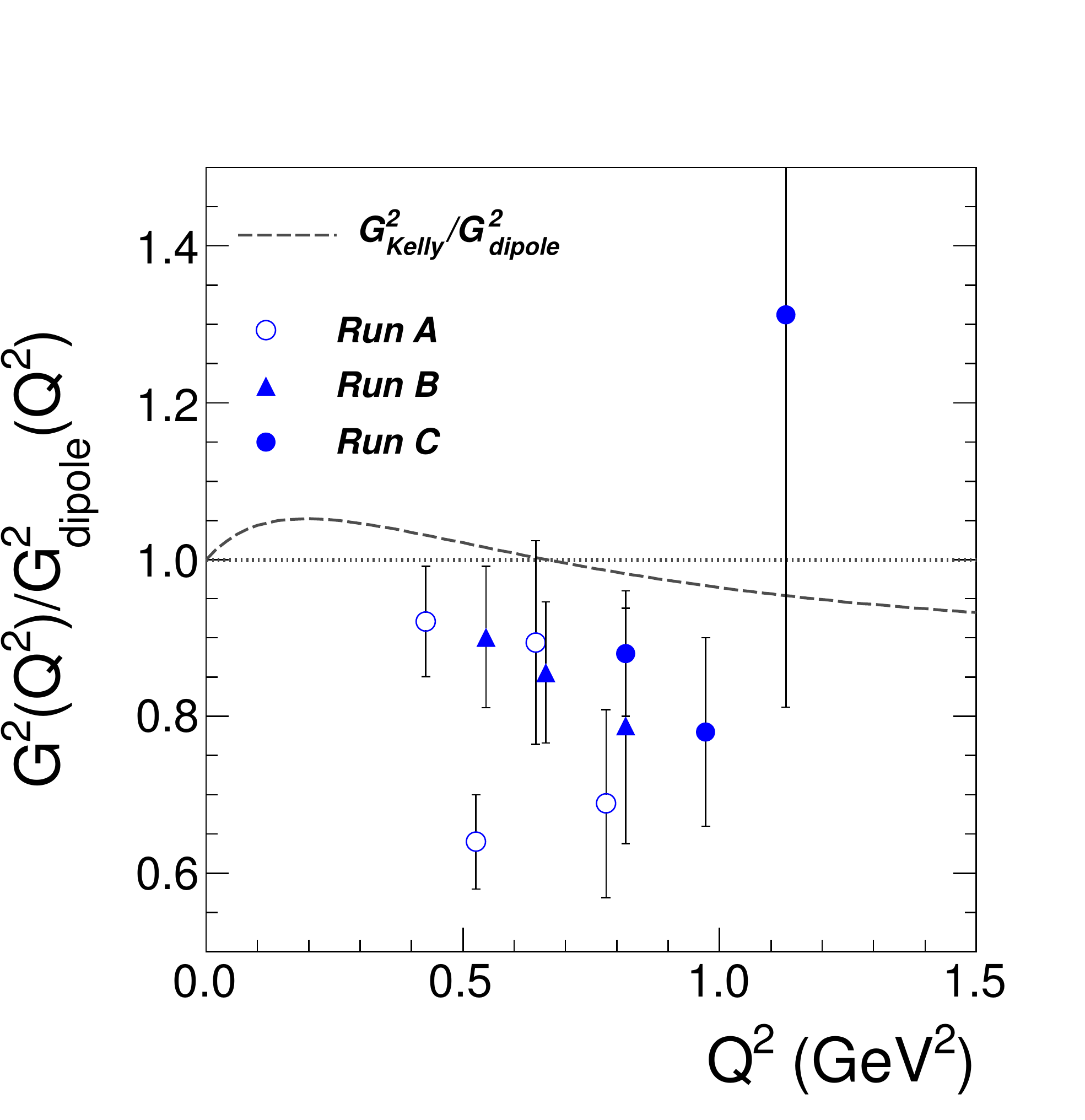}}
\caption{Ratio of the reduced cross sections,
$G^2 = d\sigma/d\Omega/d\sigma/d\Omega_{Mott}$, 
to the expected result for dipole form factors
for $\mu p$ elastic scattering,
from Ellsworth {\it et al.} \cite{Ellsworth:1968zz}.
The data are somewhat below expectations from the dipole form
factor parameterization.
Use of the more modern Kelly parameterization \cite{Kelly:2004hm}
does not qualitatively change the result.} 
\label{fig:ellsworth}
\end{figure}

One of the better early $\mu p$ elastic scattering
experiments was Ellsworth {\it et al.} \cite{Ellsworth:1968zz}, which
found that cross sections in the range $Q^2$ $\approx$ 0.5 - 1 GeV$^2$
were about 15\% below the standard dipole parameterization,
$G_E = G_M/\mu_p = (1+Q^2/0.71)^{-2}$ with $Q^2$ in GeV$^2$,
and a similar percentage below modern form factor fits,
as shown in Fig.~\ref{fig:ellsworth}.
While this suggests an $ep$ vs.\ $\mu p$ interaction difference,
Ellsworth {\it et al.}  interpreted the difference as an upper limit
on any difference in $\mu p$ and $ep$ interactions. These data are 
too high in $Q^2$ to make any inferences about the proton radius.
A subsequent experiment \cite{Camilleri:1969ag} covering 0.15 $<$ $Q^2$
$<$ 0.85 GeV$^2$ found $\mu p$ cross sections about 8\% smaller than the
electron scattering results, similar to \cite{Ellsworth:1968zz}, and
considered the $\mu p$ and $ep$ scattering results consistent within
uncertainties.
A final elastic scattering experiment \cite{Kostoulas:1973py} analyzed
the ratio of proton elastic form factors determined in $\mu p$ and $ep$
scattering as $G^2_{\mu p}/G^2_{e p} = N(1+Q^2/\Lambda^2)^{-2}$,
with the result that the normalizations are consistent with unity at
the level of 10\%, and the combined world $\mu p$ data give 
$1/\Lambda^2$ = 0.051 $\pm$ 0.024 GeV$^{-2}$, about 
2.1$\sigma$ from the electron-muon universality expectation of 0.
For deep-inelastic scattering \cite{Entenberg:1973pw}, a similar analysis
yields a normalization consistent with unity at the level of 4\%
and $1/\Lambda^2$ = 0.006 $\pm$ 0.016 GeV$^{-2}$.
In summary, old comparisons of $ep$ and $\mu p$ elastic scattering were 
interpreted as indicating no differences between $\mu p$ and 
$ep$ scattering, within the 5\% -- 10\% uncertainties of the
experiments. 
In light of the Proton Radius Puzzle, it seems that the directly measured constraints
on differing $\mu p$ and $ep$ interactions are insufficient. 
While $ep$ studies have advanced significantly in the past decades, 
the $\mu p$ work has not.

Two-photon exchange effects have also been tested in $\mu p$ scattering.
In \cite{Camilleri:1969ah}, no evidence was found for
2$\gamma$ effects, as $\mu^+p$ vs. $\mu^-p$ elastic 
scattering cross section asymmetries were consistent with 0, with 
uncertainties from 4 $\to$ 30\%, and with no visible nonlinearities
in Rosenbluth separations at $Q^2$ $\approx$ 0.3 GeV$^2$.
The Rosenbluth cross sections were determined to about 4\%.
Tests in $ep$ scattering \cite{Tvaskis:2005ex} have found no nonlinearities even with
$\approx$1\% cross sections; improved experiments are underway \cite{arringtonsr2}.
Current best estimates of the size of the nonlinearities in Rosenbluth
separations for $ep$ scattering are typically at the percent level.
Thus, it seems again in light of current knowledge that two-photon exchange 
has not been precisely enough studied in the case of $\mu p$ scattering.

The radius of $^{12}$C is one of the most precisely determined radii 
from electron scattering. The electron scattering result
\cite{Cardman:1980}
is $\langle r^2 \rangle ^{1/2}$ = 2.472 $\pm$ 0.015 fm, based
on scattering of 25 -- 115 MeV electrons at momentum transfers
from 0.1 -- 1.0 fm$^{-1}$, or $Q^2$ $\approx$ 0.0004 - 0.04 GeV$^2$.
A subsequent analysis of world data \cite{Offermann:1991ft} found that
dispersive corrections increase the extracted radius to  2.478 $\pm$ 0.009 fm.
The charge radius was also measured by determining the $\approx$90 keV 
X-ray energies in muonic carbon atoms to several eV \cite{Schaller:1982}.
Assuming a harmonic oscillator nuclear charge distribution led to
a $^{12}$C radius of $\langle r^2 \rangle ^{1/2}$ = 2.4715 $\pm$ 0.016
fm. A subsequent muonic atom experiment\cite{Ruckstuhl:1984as} found 
$\langle r^2 \rangle ^{1/2}$ = 2.483 $\pm$ 0.002 fm. 
There is a consistent result for the carbon radius from a 
$\mu$C scattering experiment \cite{PhysRevC.8.896}, but with uncertainties an order of magnitude worse. 
There is evidently no $\mu p$ vs.\ $ep$ issue in the carbon radius
determination.
There are several possible reasons why there might be a $\mu$ / $e$
difference in the proton but not in carbon.
Examples include opposite effects in the case of $\mu n$ vs.\ $\mu p$
interactions, and the charge distribution in carbon resulting largely
from orbital motion of the nucleons, in which there is no effect,
vs. charge distributions of the nucleons, in which there is an effect.

To summarize, direct comparisons of $\mu p$ and $ep$ scattering
were done, but with poor overall precision.
The comparisons were also at sufficiently large $Q^2$ that they
would not be sensitive to the proton radius.
Measurements sensitive to 2$\gamma$ exchange were also performed, but
at a level that we now believe is not sufficiently precise to
provide significant results.
While the carbon radius is much better determined, and is consistent
for muon and electron measurements, the implications of this for the
Proton Radius Puzzle are not clear.

\subsection{Motivation Summary}

The Proton Radius Puzzle has attracted wide interest,
but the resolution to the Puzzle is unclear.
It might arise from beyond standard
model physics, novel hadronic physics, 
or issues and / or underestimated uncertainties in the 
determination of the radius from the actual experimental data.
There is strong support in the community for a number of
experiments that test different explanations for the Puzzle.
New $ep$ atomic physics and scattering experiments are planned,
as are additional muonic atom experiments.

The MUSE experiment presented here is the only proposed
$\mu p$ elastic scattering experiment.
MUSE intends to
\begin{itemize}
\item measure both $\mu p$ and $ep$ scattering in the low $Q^2$ region,
\item measure both charge signs,
\item extract form factors and proton radii,
\item compare $ep$ and $\mu p$ scattering, form factors, and radii
   as a test of lepton (non-)universality,
\item study the possibility of unexpected structures and/or
  extrapolation errors affecting the radius extraction, and
\item determine two-photon exchange effects, to test their effect on
  the radius extraction and to test possible hadronic physics
  explanations of the Puzzle.
\end{itemize}
Thus the MUSE experiment looks at several possible explanations of the
Proton Radius Puzzle.

\section{Measurement Overview}

The MUSE measurement is planned for the $\pi$M1 beam line at
the Paul Scherrer Institut (PSI), in Villigen Switzerland.
The MUSE approach to resolving the Proton
Radius Puzzle is to measure simultaneously elastic 
$\mu^{\pm} p$ scattering and $e^{\pm} p$ scattering.
The $\mu p$ scattering will be compared to $ep$ scattering at the 
cross section level, with extracted form factors, and
ultimately with an extracted radius.
Measurements with the two beam polarities will be compared to
determine the (real part of the) two-photon exchange.
The basic idea is to provide a higher precision comparison of 
$\mu p$ and $ep$ interactions in a region sensitive to the proton
radius, and to check that the two-photon exchange is under control,
and does not distort the extraction of the radius.
At the same time, these data can check predictions of enhanced
two-photon exchange from novel hadronic physics, and certain
BSM physics models that affect the form factors in $\mu p$ vs.\
$ep$ determinations.

In electron scattering, high precision experiments have typically
used an intense, low-emittance beam incident on a cryotarget,
with scattered particles detected by a high-resolution, small solid angle spectrometer.
A muon scattering experiment must be different because the
intense low-emittance primary electron beam is replaced by an 8 -- 9 orders
of magnitude less intense, much larger emittance, secondary muon beam, which is
also contaminated with electrons and pions.
To run a high precision experiment in these conditions requires
several adjustments.
The low intensity necessitates a large acceptance spectrometer and long run times.
The large emittance necessitates measuring the individual beam particle incident
trajectories.
The presence of several different particle species in the beam
requires identifying each individual beam particle type.

The difficulties of muon scattering are in part compensated by several
advantages.
Since the muon beam is a secondary beam, one can easily obtain
essentially identical beams of both charge signs, which allows a precise
determination of two-photon exchange effects.
Conventional two-photon effects are expected to be of order 1\% --
though they have not been measured that precisely -- and
have the potential to affect the extracted radius; there is also the
possibility that the proton polarizibility is the underlying cause of
the Puzzle, and it will lead to enhanced two-photon exchange.
Here the effects of two-photon exchange can be
determined and the average of $\mu^{\pm} p$ cross sections removes
the two-photon exchange contributions from the cross sections and the
form factors.
The low muon intensity eliminates target density fluctuations from
beam heating.
The electron contamination in the beam allows a simultaneous
measurement of $e p$ scattering for comparison with the muon scattering.
The use of a non-magnetic spectrometer allows the solid angle to be
determined more precisely than is typically possible with a magnetic
spectrometer.

A precise measurement also requires an amount of kinematic overlap,
measuring cross sections multiple times to ensure that the 
experimental systematics are well understood.
In electron scattering experiments this can be done with
multiple beam energies and overlapping spectrometer settings, using
a monitor spectrometer to confirm the relative luminosity for each
setting at a fixed beam energy.
In MUSE the overlap is provided by using 3 beam momenta and two
independent large solid angle spectrometer systems.
A run with the spectrometer wire chambers rotated
by a small angle is also planned as a cross check.

MUSE runs in several stages.
Initial beam tests in Fall 2012 \cite{MUSEtest:2012} verified the basic properties
of the muon beam in the $\pi$M1 beam line at PSI.
A second round of beam tests will run in summer
2013; these tests will study beam properties in more
detail using GEM chambers, prototype a quartz Cherenkov
detector, and do a simplified scattering experiment to verify
simulated backgrounds.
As equipment is constructed, we expect additional beam
tests of various experiment components, described in more detail
below, leading up to a two-month ``dress rehearsal'' measurement with
beam line detectors and at least one spectrometer, perhaps
in late 2015.
The dress rehearsal is intended to be a high statistics study to
investigate any potential issues with the equipment as built or with backgrounds.
Assuming analysis of this initial high statistics measurement confirms
the experiment functionality, MUSE is ready 
to commence a two year production run.

\section{Experimental Details}

Here we present a summary of the MUSE technical design;
more details can be found in \cite{MUSETDR:2013}

\subsection{Muon Beam Line}
\label{sect:muonbeam}
The PSI $\pi$M1 beam line provides
a mixed muon / pion / electron beam with a $\approx$50 MHz time structure.
The three beam momenta selected, 
$p_{in}$ $\approx$ 115 MeV/$c$, 153 MeV/$c$, and 
210  MeV/$c$, are chosen both to cover a kinematic range
and provide overlaps, and because at these three momenta,
with the expected detector geometry,
the different beam particle types can be efficiently separated
using RF time measurements.
Magnet polarities can be reversed to allow the channel to 
transport either positive or negative polarity particles. 

\subsection{Detector Overview}
\label{sec:det_system}

\begin{figure}[hb]
\centerline{\includegraphics[width=2.5in]{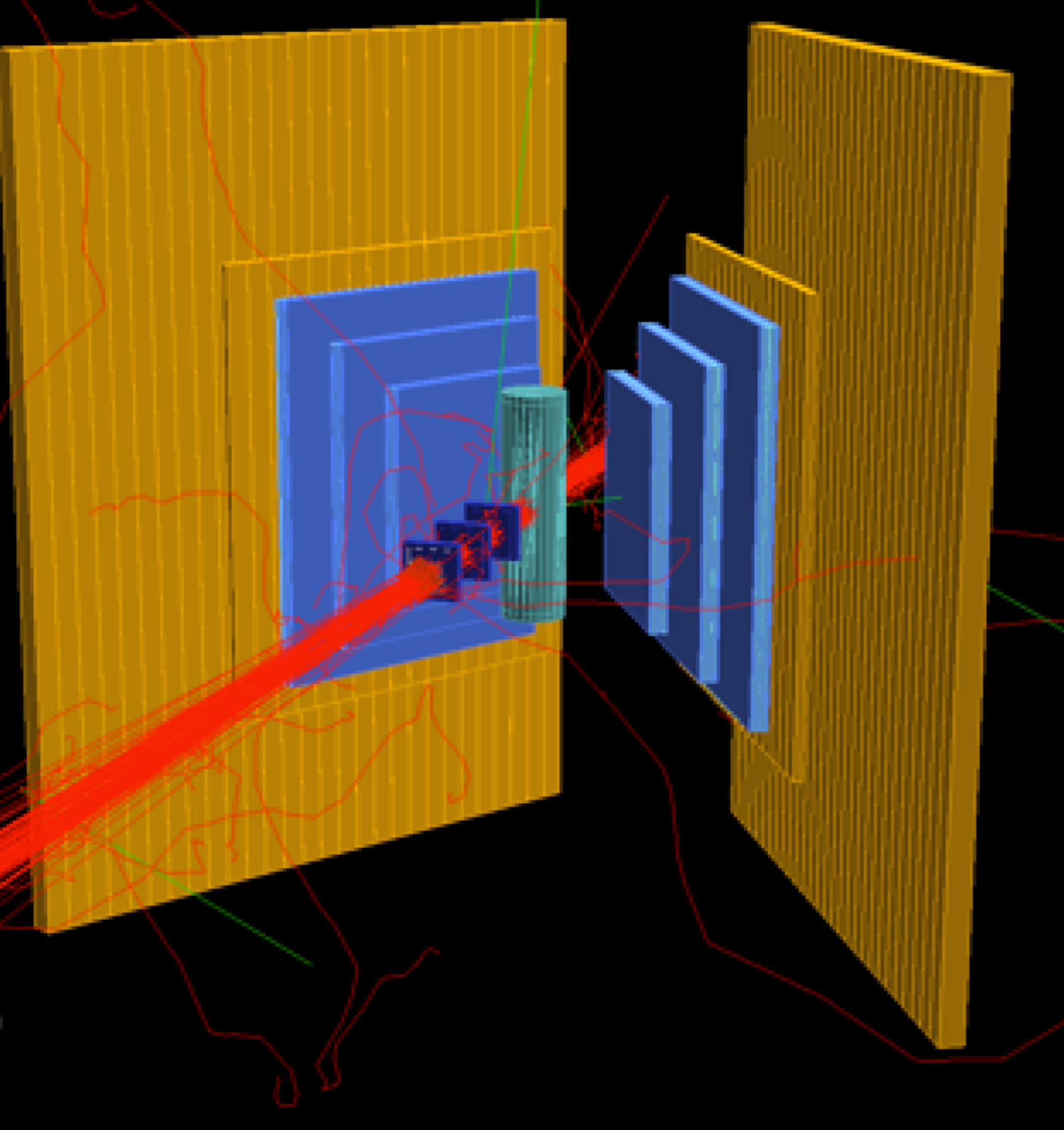}}
\caption{A Geant4 simulation showing part of the MUSE
  experimental system. Here one sees the beam going through the
GEM chambers and the scattering chamber, along with the spectrometer
wire chambers and scintillator hodoscopes.
The beam SciFi's, quartz Cherenkov, and beam monitor scintillators are
missing from this view.
} 
\label{fig:detectcartoon}
\end{figure}

The $\pi$M1 channel features a momentum dispersed ($\approx$7 cm/\%)
intermediate focal point (IFP) and a small beam spot ($\sigma_{x,y}$ $<$ 1 cm) 
at the scattering target.
The base line design for the MUSE beam detectors has a collimator and
a scintillating fiber detector (SciFi) at the intermediate focus.
Some of the detectors in the target region are shown in Fig.~\ref{fig:detectcartoon}.
After the channel and immediately before the target there are
a SciFi detector, a quartz Cherenkov detector, and a set of GEM
chambers.
A high precision beam line monitor scintillator hodoscope
is downstream of the target.

The IFP collimator serves to cut the $\pi$M1 channel acceptance
to reduce the beam flux to manageable levels.
The IFP SciFi measures the RF time, for use in determining particle
type, and measures beam particle position, to determine the particle
momentum and thus the beam momentum spectrum.

The target SciFi measures the RF time, for use in determining
particle type.
The quartz Cherenkov provides higher resolution timing which will be
used at the analysis level to reject muon decay events.
The GEM chambers determine the trajectory of particles incident upon
the target.
The time at the target, in conjunction with the time at the IFP,
provides a time of flight measurement of the beam particles over
a path length of about 9 m, providing additional identification
capability at the analysis level.

The beam line monitor hodoscope is intended to provide a high
resolution determination of the RF time for randomly coincident
unscattered beam particles. This monitors the stability of the 
channel and timing with the accelerator RF signals.

The base line design for the scattered particle spectrometers is
a set of three wire chambers followed by two scintillator hodoscopes.
The wire chambers provide outgoing trajectories, that are used
in combination with the tracks found by the GEM chambers 
to determine scattering angles and interaction positions.
The two scintillator hodoscopes provide high resolution timing,
high efficiency triggering, and limited position information.

The detector systems operate in a triggered mode with
VME-based readout.
The much larger pion scattering cross section necessitates
using custom field programmable gate array (FPGA) units to 
determine beam particle type from the SciFi signals at the hardware level.
The trigger is based on this hardware determination
of beam particle type along with a trigger matrix (also implemented
in an FPGA) for the scattered particle scintillator hodoscopes to limit
triggers to events with trajectories pointing approximately 
to the target.
VME modules are a mix of mostly time and charge to digital
converters (TDCs and QDCs).
The MIDAS data acquisition system developed at PSI by Stefan
Ritt was used in the initial test run and is planned to be used in
MUSE.

The hardware components of MUSE are largely established technology.
SciFi detectors are now common.
The use of a quartz Cherenkov detector to provide $\approx$10 ps
timing has been prototyped by a group at Fermilab \cite{Albrow:2012ha}.
The GEM chambers exist already, having been used in the OLYMPUS
experiment at DESY.
The high precision scintillators, used both in the spectrometer and for beam
line monitoring, copy a design already constructed and
tested for the Jefferson Lab CLAS 12 upgrade.
The wire chamber designs are based upon the chambers built at
University of Virginia for the Hall A Bigbite Spectrometer.

For the trigger and readout electronics, a mixture of existing
commercial equipment and custom or recently prototyped
boards is planned.
The beam particle identification system will be implemented
in commercial FPGAs, but the FPGAs will be installed in custom
designed boards.
The trigger uses a CAEN FPGA.
We note here that the use of FPGAs in subatomic physics experiments
has become fairly commonplace.
To contain costs, time measurements will be done with the 
recently prototyped TRB3 TDC,
developed GSI in Darmstadt, Germany.

\subsection{Cryotarget}
\label{Target}

Liquid hydrogen targets in vacuum systems are a mature
technology.
The design will follow standard, well known
and tested cryogenic cell designs.
The MUSE target is a relatively easy target, as the beam power deposited
in the target is a few $\mu$W.
The main concerns then are residual air in the vacuum system freezing to
the target, and radiative heating of the target by the vacuum system;
both issues can be ameliorated through liquid nitrogen baffles in the
scattering chamber.
The base line design is for the cryotarget system to have
a target ladder containing the cryogenic cell, constructed from thin
kapton,
a dummy target for wall backgrounds,
a carbon target for positioning,
and an empty target position.

\section{Collaboration Responsibilities and Commitments  from PSI}

The core of the MUSE collaboration is the
institutions making a commitment to develop major parts of the
experiment and/or have Ph.D. students and
postdocs essentially fully committed to the experiment.
A summary of commitments to the basic equipment development
and some other tasks is shown in Table~\ref{tab:musesplit}.
Several of the institutions -- GW, Hebrew University, MIT, Rutgers,
and Tel Aviv -- have committed to having Ph.D.\ students and / or 
postdocs spend significant fractions of their time at PSI for the experiment.

\begin{table}[h]
\caption{\label{tab:musesplit} MUSE equipment responsibilities.}
\begin{tabular}{|c|c|c|}
\hline 
\textbf{Device} & \textbf{Institution} & \textbf{Person}\tabularnewline
\hline 
\hline 
$\pi$M1 Channel & PSI & K. Dieters  \tabularnewline
\hline 
Scintillating Fibers & Tel Aviv &  E. Piasetzky \tabularnewline
\hline
Scintillating Fibers & St. Mary's & A. Sarty \tabularnewline
\hline 
GEM chambers (existing) & Hampton & M. Kohl  \tabularnewline
\hline
Beam Quartz Cherenkov & Hebrew University & G. Ron (Co-Spokesperson) \tabularnewline
\hline
Cryogenic Target System & Hebrew University & G. Ron (Co-Spokesperson) \tabularnewline
\hline 
Wire Chambers & M.I.T. & S. Gilad \tabularnewline
\hline 
Scintillators &  South Carolina & S. Strauch \tabularnewline
\hline 
Electronics and Trigger & Rutgers & R. Gilman  (Spokesperson)\tabularnewline
\hline 
Readout Electronics and DAQ System & George Washington & E. J. Downie (Co-Spokesperson)\tabularnewline
\hline 
Data Acquisition Software & MIT \& Rutgers & V. Sulkosky \& K. Myers \tabularnewline
\hline 
Radiative Corrections & George Washington & A. Afanasev \tabularnewline
\hline 
Analysis and Radius Extraction & Argonne & J. Arrington \tabularnewline
\hline 
\end{tabular}
\end{table}

\subsection{Schedule}

MUSE \cite{MUSEPROP:2013} was approved by the PSI PAC in Jan 2013.
A second test run is planned for summer 2013.
It is the intent of the collaboration to seek funding
during 2013, so that equipment construction can start in 2014.

Construction of the experiment requires about two years.
To a large degree, the beam detectors are all small and can be
constructed in several months. The time needed
for procurement and testing will result in these detectors
being available in about 9 - 12 months after funds are
available.

The cryotarget, high precision scintillators, trigger, and 
wire chambers require more time.

The cryotarget requires about 2 years to construct. Designing the
target, purchasing components, and assembling the basic system 
requires about 12 months. Installing and commissioning the control
system will require an additional 9 months. At this point the target
can be cooled and tested, which requires an additional 3 months.

The high-precision scintillators are similar to those constructed 
at South Carolina for the CLAS 12 GeV upgrade. The exact construction
rate depends on the number and expertise of students involved in building and
testing the scintillators; we expect the average 
production rate to be about two scintillator paddles per week.
Production will come up to speed faster if experienced students
from the CLAS 12 project are still available.
In addition to the 1 year needed to build the scintillators,
an additional 6 months will be needed for procurement, testing, 
and shipping.
Thus, the entire scintillator project will require about 18 months.
It should be possible to start the initial procurement activities,
such as obtaining bids, before funds arrive.
As a result, it should be possible to have all the scintillators
needed for one spectrometer for a dress rehearsal run in late 2015,
and the full complement of scintillators for production running in 2016.

Constructing the beam PID requires design work, prototyping, extensive
programming, and design and construction of the final system.  
FPGAs often exhibit quirky and interesting behavior.
Thus, even when the FPGA selected for the project is chosen
appropriately, and even though the estimate of the time for 
the project comes from an experienced FPGAs programmer, 
the project time can exceed estimates. The Rutgers electronics
shop also has LHC projects that will compete for programmer time.
An initial system should be ready for the dress rehearsal, in just
over a year, and the experience gained should allow the full system
to be deployed within 2 years, in time for the production data.
The trigger FPGA system can be developed in parallel with the beam
PID FPGA system; it is a simpler programming challenge that uses
commercial equipment, so it should be ready sooner.
This part of the system can be developed by students and postdocs.

The wire chambers are the most time-consuming construction project.
It requires about 6 months of design, procurement, and preparation
before wire chamber construction can begin.
We assume here that clean room space can be found, so that a new
clean room does not need to be constructed.
Initially, as the chamber workers are trained, it will take about 3
months to produce the first wire chamber, and an additional month
to test it. Each subsequent chamber can be produced in slightly
less time, but it will require about two years to produce all the
chambers. Within a year it should be possible to have 
available at least two chambers for one of the spectrometers 
for a dress rehearsal run.
With sufficient space and personnel, it might be possible to produce
two chambers in parallel and shorten the production time.

The equipment, on being brought to PSI, has to be installed, hooked up
to electronics, etc., and commissioned. Doing this for the entire set
of experimental apparatus will take about 6 months, but as indicated
above the equipment is expected to arrive over a period of about 1 year.

Assuming that PSI continues to run during the second half of each
calendar year, and assuming that funding for equipment construction
is received in early - mid 2014, it should be possible to have a
significant fraction of the MUSE equipment on hand for a significant
test of the system in late 2015. The beam line detectors,
scintillators, some of the wire chambers, and a simplified version
of the trigger should all be available. The cryotarget will not be ready,
but solid targets can be used for initial testing.

\subsection{PSI Commitments}

The verbal close out of the January 2013 PSI PAC
concerning the MUSE experiment was:
``We are certainly convinced that the proton radius puzzle is an important physics puzzle,
largely this lab is responsible for that, and therefore it is totally fitting to finding a solution
to it. So we approve the experiment, we want to see it done. We are very pleased by the
progress made last year in the beam test, a lot of lessons were learned, a few things were
not quite as optimistic as hoped, on the other hand there is nothing there which was a
major problem.''

PSI was an excellent host for our test beam time in 2012.
We were provided with access to $\pi$M1, beam time,
installation assistance,
office space, access to infrastructure such as computer networking,
and the use of large amounts of existing experimental 
equipment, such as electronics.

PSI will be providing us with additional beam time in 2013,
along with similar access to that which we had in 2012.
The laboratory is making minor adjustments to the $\pi$M1 channel
for our tests: installation of an NMR to monitor dipole stability,
installation of a collimator at the intermediate focus, 
and adjustments to quadrupoles to fine tune the positioning of the 
beam focus.
Also planned for the future are minor adjustments to vacuum pipes 
in the downstream half of the beam line, and the possible addition of a concrete
shielding wall just before the detectors.

\section{Future Plans}

The equipment to be constructed for this experiment is versatile 
enough to be used as part of several measurements at 
PSI, as well as potential future measurements at US and other
worldwide facilities. 

Depending on the results of MUSE and other Proton Radius Puzzle
experiments, there are natural follow
up $\mu p$ scattering measurements to be performed.
One is a measurement of enhanced parity violation as predicted by
certain BSM models.
A second would be a higher precision measurement focused on the
two-photon exchange contributions at large angles.
A third would be to move the apparatus to a different PSI muon beam line to 
obtain lower momentum surface muons, to reach even lower beam momenta
and momentum transfer.

Another direction is determining the radii of light nuclei with muon
scattering.
The PSI CREMA collaboration, responsible
for the muonic hydrogen measurements, intends to measure nuclei such as
$^3$He, $^4$He, $^6$Li, and $^{11}$B.
Some have recently been measured at JLab to high precision.
Muon scattering can determine the radii of these nuclei or others,
such as $^{12}$C, which was already measured, but
with low precision.
Of particular interest is a measurement on deuterium, 
which will also allow the only extraction of the muonic neutron
radius. 
Additionally, some US groups have expressed 
interest in extending the measurements to include $^3$H charge radii. 

\section{Summary}

The Proton Radius Puzzle is arguably the most pertinent, controversial 
and timely issue in the Hadron Physics community at this present time.  
The discrepancy between the proton charge radius as measured with 
muons and that measured in electron experiments, in both scattering 
and excitation spectra-based extractions, is widely recognized and needs to 
be explained, as stated in the CODATA analysis and the Particle Data
Group review, and reiterated by the JLab and PSI PACs.
No resolution to the Puzzle has been found, and it has attracted
widespread interest.

The MUSE experiment 
measures muon- and electon-proton elastic scattering, 
at the same time with the same equipment, which will allow:
\begin{itemize}
\item  The highest precision scattering experiment
determination of the consistency of the $\mu p$ interaction with the
$ep$ interaction, through cross sections and extracted form factors 
and radii.
\item A test of the importance of 2$\gamma$ exchange effects.
\item Checks of possible explanations of the puzzle including
structures in the form factors, extrapolation errors in the radius
extraction from scattering measurements, anomalously large two-photon
effects leading to issues in extracting the radius, including possible
effects from proton polarizibility, and possible electron-muon differences. 
\end{itemize}
MUSE  provides the missing measurement of the four possible radius
determinations using scattering or atomic energy levels of  $\mu p$
and $ep$ systems, and tests several possible explanations of the 
Proton Radius Puzzle.
The experiment is technically feasible
on a time scale of about 4 years.

We acknowledge the support of the US Department of Energy,
the US National Science Foundation, and the Paul Scherrer Institut,
in the early stages of the MUSE experiment.

\end{spacing}

\bibliography{muse_WP_v2}

\end{document}